\documentclass[prl,aps,twocolumn,superscriptaddress,showpacs]{revtex4}
\usepackage{amsmath,amssymb,graphicx}
\usepackage{pxfonts}
\usepackage{graphicx}
\usepackage{color}
\usepackage{float}

\begin{document}

\date{\today}

\title{Interference dislocations in condensate of indirect excitons}

\author{J.\,R.~Leonard}, \author{Lunhui~Hu}, \author{A.\,A.~High}, \author{A.\,T.~Hammack}, \author{Congjun~Wu}, \author{L.\,V.~Butov}
\affiliation{Department of Physics, University of California at San Diego, La Jolla, California 92093-0319, USA}
\author{K.\,L.~Campman}, \author{A.\,C.~Gossard}
\affiliation{Materials Department, University of California at Santa Barbara, Santa Barbara, California 93106-5050, USA}


\date{\today}

\begin{abstract}
\noindent
Phase singularities in quantum states play a significant role both in the state properties and in the transition between the states. For instance, a transition to two-dimensional superfluid state is governed by pairing of vortices and, in turn, unpaired vortices can cause dissipations for particle fluxes. Vortices and other phase defects can be revealed by characteristic features in interference patterns produced by the quantum system. We present dislocation-like phase singularities in interference patterns in a condensate of indirect excitons measured by shift-interferometry. We show that the observed dislocations in interference patterns are not associated with conventional phase defects: neither with vortices, nor with polarization vortices, nor with half-vortices, nor with skyrmions, nor with half-skyrmions. We present the origin of these new phase singularities in condensate interference patterns: the observed interference dislocations originate from converging of the condensate matter waves propagating from different sources.
\end{abstract}
\maketitle

Phase singularities are studied in quantum states of matter ranging from superconductors and superfluid Helium~\cite{Volovik2003} to condensates of atoms~\cite{Matthews1999, Inouye2001, Hadzibabic2006, Khawaja2001, Kamchatnov2008, Wu2008, Wang2010}, magnons~\cite{Nowik-Boltyk2012}, polaritons~\cite{Rubo2007, Lagoudakis2008, Lagoudakis2009, Roumpos2011, Amo2011, Grosso2011, Flayac2011, Hivet2012, Flayac2013, Bardyn2015}, and excitons~\cite{Wu2008, High2012, Matuszewski2012, Kyriienko2012, High2013, Kavokin2013, Vishnevsky2013, Sigurdsson2014, Bardyn2015, Li2016, Leonard2018}. A variety of phase defects is considered, including vortices~\cite{Matthews1999, Inouye2001, Hadzibabic2006, Lagoudakis2008, Roumpos2011, Nowik-Boltyk2012}, polarization vortices~\cite{High2012, High2013, Kavokin2013}, half-vortices~\cite{Volovik2003, Rubo2007, Wu2008, Lagoudakis2009, Sigurdsson2014}, skyrmions~\cite{Khawaja2001, Wu2008, Vishnevsky2013, Flayac2013, Li2016}, solitons~\cite{Kamchatnov2008, Amo2011, Grosso2011, Flayac2011, Hivet2012, Flayac2013}, striped phases~\cite{Wang2010, Matuszewski2012, Kyriienko2012, Vishnevsky2013, Kavokin2013, Sigurdsson2014}, and phase domains~\cite{Leonard2018}.

\begin{figure}
\begin{center}
\includegraphics[width=5.3cm]{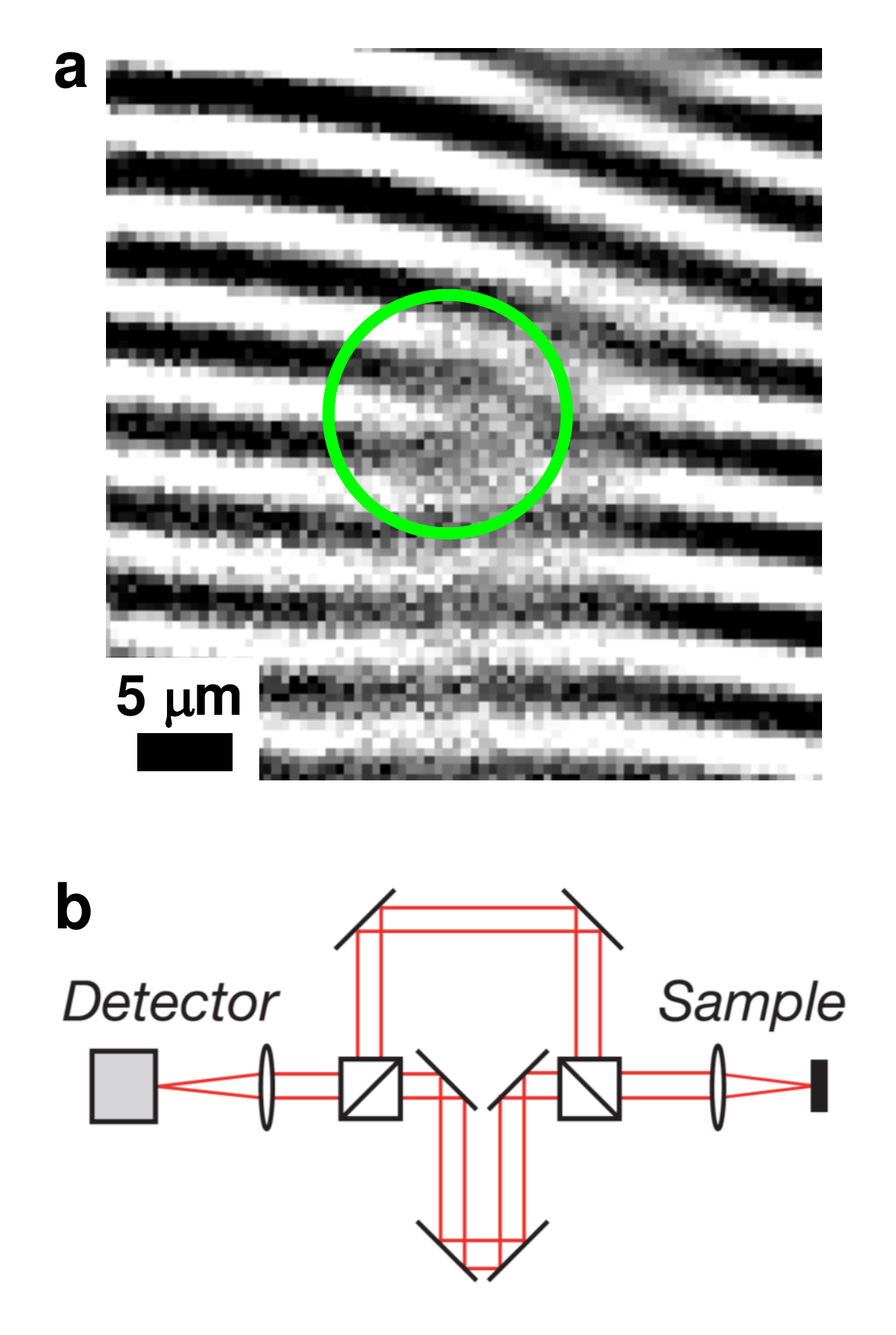}
\caption{\textbf{Dislocation-like singularity in the exciton interference pattern.} (a) Measured interference pattern $I_{\rm interf}(x, y)$. The dislocation in the interference pattern is marked by a green circle. (b) Diagram of the interferometric set-up. The emission images produced by each of the two arms of the Mach-Zehnder interferometer are shifted with respect to each other in the CQW plane.
}
\end{center}
\label{fig:spectra}
\end{figure}

Quantum states of excitons can be created in a system of indirect excitons (IXs), aka interlayer excitons~\cite{Lozovik1976}. IXs are formed by electrons and holes confined in spatially separated layers in coupled quantum wells (CQW). Due to their long lifetimes IXs can cool below the temperature of quantum degeneracy and form a condensate in momentum space~\cite{High2012}. The IX condensation is detected by the measurement of IX spontaneous coherence with a coherence length much larger than in a classical gas~\cite{High2012}.

A cold gas of IXs is realized in regions of external ring and localized bright spot (LBS) rings in IX emission~\cite{High2012}. These rings form on the boundaries of electron-rich and hole-rich regions created by current through the structure and optical excitation, respectively; see Ref.~\cite{Yang2015} and references therein. LBS sources are stable and well defined sources of cold IXs~\cite{Yang2015} thus a suitable system for exploring phenomena in exciton condensates.

\begin{figure*}
\begin{center}
\includegraphics[width=17.7cm]{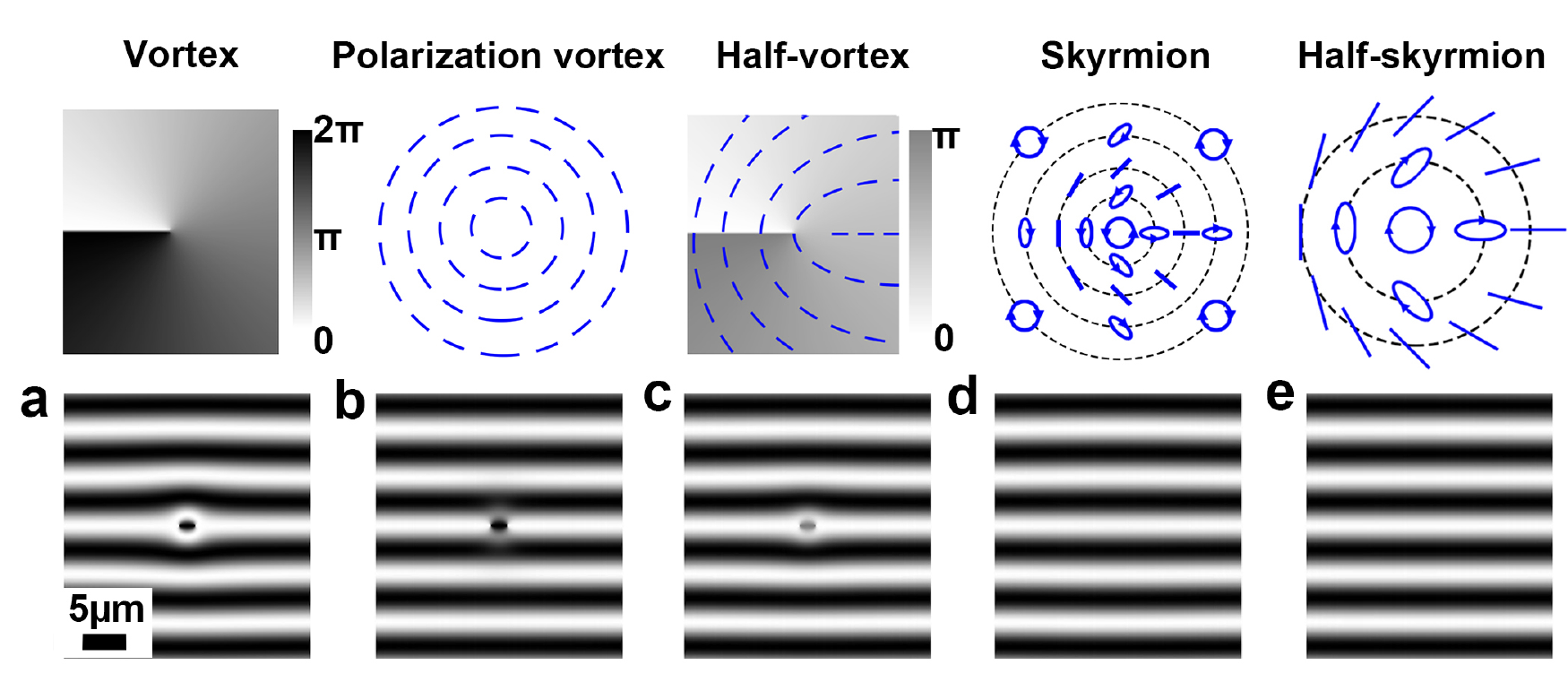}
\caption{\textbf{Simulated exciton interference patterns for a vortex, polarization vortex, half-vortex, skyrmion, and half-skyrmion.} Interference patterns $I_{\rm interf}(x, y)$ for a phase vortex (a), polarization vortex (b), half-vortex (c), skyrmion (d), and half-skyrmion (e) for the shift-interferometry corresponding to the experiment (Fig.~1). The phase and polarization patterns are shown on the top: the phase is presented by the color and the linear, circular, and elliptic polarizations are presented by bars, circles, and ellipses, respectively. None of these simulated interference patterns is similar to the experiment (Fig.~1a), indicating that the observed singularity in the interference pattern (Fig.~1a) is not associated with a vortex, or polarization vortice, or half-vortice, or skyrmion, or half-skyrmion.
}
\end{center}
\label{fig:spectra}
\end{figure*}

The earlier studies~\cite{High2012} demonstrated dislocation-like (fork-like) singularities in interference patterns produced by IX condensate. However, the origin of these phase singularities remained unknown. Here, we show that the observed dislocations in interference patterns are not associated with conventional phase defects: neither with vortices, nor with polarization vortices, nor with half-vortices, nor with skyrmions, nor with half-skyrmions. We present the origin of these new singularities: we show that the observed interference dislocations originate from converging of the condensate matter waves propagating from different sources.

Figure~1a shows the interference pattern of IX emission measured by shift-interferometry. The emission images produced by each of the two arms of the Mach-Zehnder interferometer (Fig.~1b) are shifted with respect to each other to measure the interference between the emission of IXs separated by $\delta x$ in the CQW plane. Since excitons directly transform to photons inheriting exciton coherence, the shift-interferometry imaging of IX emission allows probing coherent phenomena in the IX system~\cite{High2012}. Figure~1a shows a dislocation-like (fork-like) phase singularity in the interference pattern. The origin of the observed dislocations in the IX interference pattern is considered below.

We start from noting that dislocations (forks) in interference patterns can be associated with vortices in quantum systems. In a singly quantized vortex, the phase of the condensate winds by $2\pi$ around the singularity point, which can be revealed as a fork-like defect in an interference pattern. Forks in interference patterns have been reported for vortices in atom condensates~\cite{Inouye2001}, polariton vortices~\cite{Lagoudakis2008, Roumpos2011, Lagoudakis2009}, magnon vortices~\cite{Nowik-Boltyk2012}, and optical vortices~\cite{Scheuer1999}.

However quantized vortices lead to the appearance of dislocations in interference patterns only for certain interferometric experiments, in particular for the interference of a vortex field with a plane wave. For our shift-interferometry experiment, simulations show that a quantized vortex should lead to the appearance of a pair of left- and right-oriented dislocations of interference fringes separated by a distance equal to the shift $\delta x$ in the shift-interferometry experiment (Fig.~2a). The shift-interference pattern simulated for a quantized vortex (Fig.~2a) is different from the experiment (Fig.~1a) indicating that the observed singularity in the interference pattern is not associated with a quantized vortex.

We also simulated the shift-interference patterns for other phase defects, including polarization vortices (Fig.~2b), half-vortices (Fig.~2c), skyrmions (Fig.~2d), and half-skyrmions (Fig.~2e). None of these simulated patterns corresponds to the experiment (Fig.~1a), indicating that the observed singularity in the interference pattern is not associated with a polarization vortice, or half-vortice, or skyrmion, or half-skyrmion.

The origin of the observed singularity in the interference pattern is outlined below. In the shift-interferometry experiment, IXs with wave function $\psi({\bf r})$ produce the interference pattern $I({\bf r}) = \vert \psi({\bf r} - \delta {\bf r} / 2) e^{iq_t y} + \psi({\bf r}+\delta {\bf r} / 2) \vert^2$, where $q_t = 2\pi\alpha/\lambda$ sets the period of the interference fringes, $\alpha$ is a small tilt angle between the image planes of the interferometer arms, and $\lambda$ is the emission wavelength.

\begin{figure}
\begin{center}
\includegraphics[width=8.7cm]{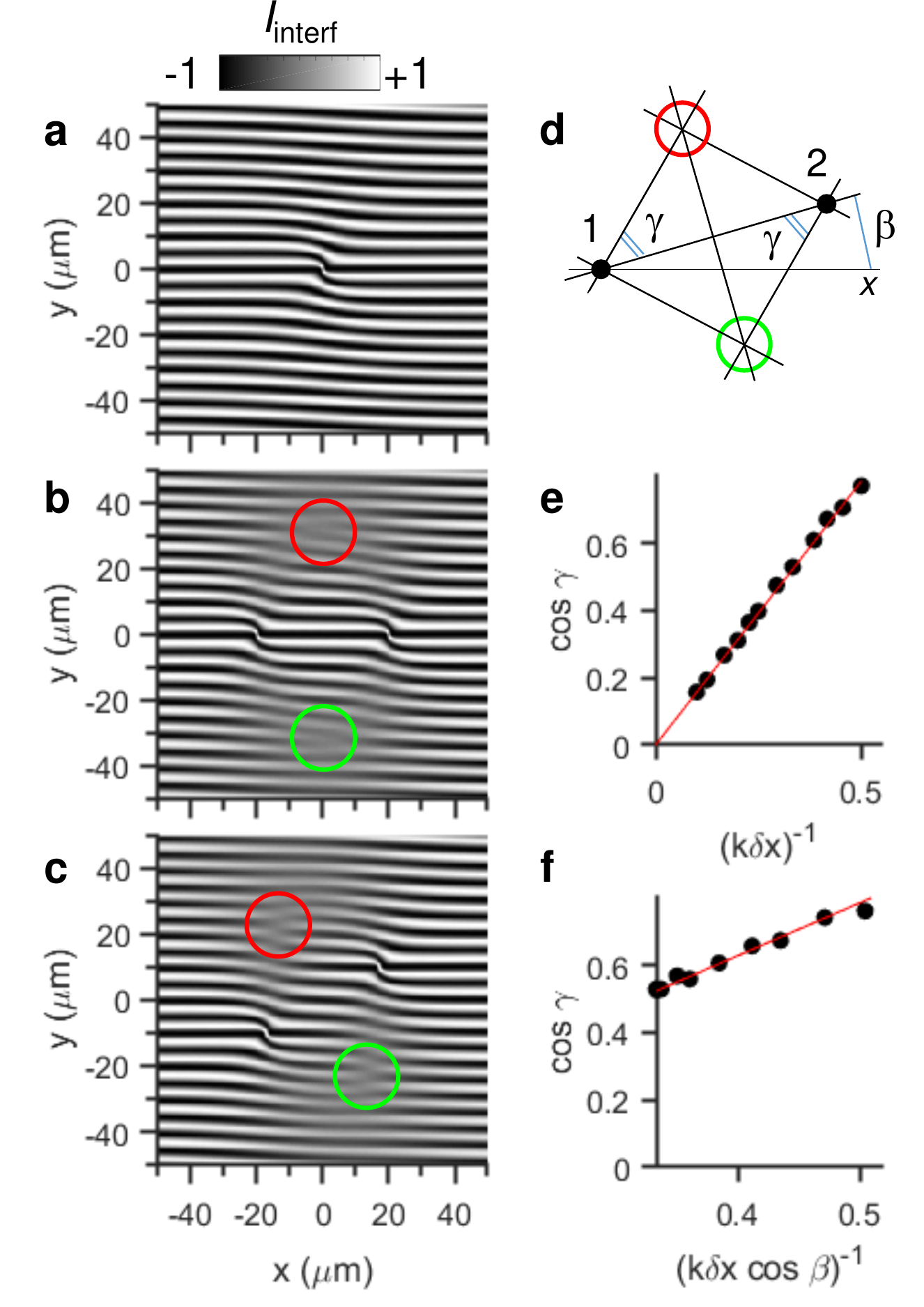}
\caption{\textbf{Simulated exciton interference patterns with the singularities.} (a) Shift-interference patterns $I_{\rm interf}(x, y)$ for IXs radially propagating from a source located in the center. $k = 1.5$~$\mu$m$^{-1}$. (b) $I_{\rm interf}(x, y)$ for IXs radially propagating from two sources separated along $\bf x$. The phase singularities in the interference pattern (marked by green and red circles) are observed at the angle $\gamma = \arccos [(k \delta x)^{-1} \pi/2]$. This relation is verified in (e) showing $\cos \gamma$ at the singularity location (points) vs $(k \delta x)^{-1}$ for varying $k$. The angle of the line is $\pi/2$. (c) $I_{\rm interf}(x, y)$ for IXs radially propagating from two sources separated along the line at the angle $\beta$ relative to $\bf x$. The phase singularities in the interference pattern (marked by green and red circles) are observed at the angle $\gamma = \arccos \left[ (k \delta x \cos \beta)^{-1} \pi/2 \right]$. This relation is varified in (f) showing $\cos \gamma$ at the singularity location (points) vs $(k \delta x \cos \beta)^{-1}$ for varying $\beta$. The angle of the line is $\pi/2$. (d) Schematics showing IX sources (1 and 2) and the singularities (green and red circles) in the interference pattern.
}
\end{center}
\label{fig:spectra}
\end{figure}

\begin{figure}
\begin{center}
\includegraphics[width=8.5cm]{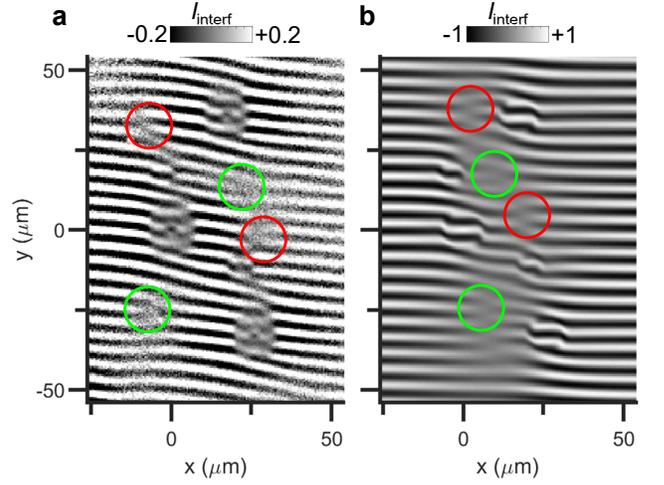}
\caption{\textbf{Measured and simulated exciton interference patterns with the singularities.} (a) Measured shift-interference patterns $I_{\rm interf}(x, y)$ for IXs in the region of five LBS sources. Three strong sources are clearly seen due to the pronounced phase domains at the source locations. These phase domains are associated with the Pancharatnam-Berry phase and are described in Ref.~\cite{Leonard2018}. Two weaker sources can also be seen $\sim 20$~$\mu$m above the bottom and medium strong sources. The right- and left-oriented dislocations in the interference pattern are marked by red and green circles, respectively. (b) Simulated $I_{\rm interf}(x, y)$ for IXs radially propagating from five sources positioned at the LBS locations. The phase domains described in Ref.~\cite{Leonard2018} are also shown. The simulations (b) qualitatively reproduce the phase singularities marked by red and green circles observed in the experiment (a).
}
\end{center}
\label{fig:spectra}
\end{figure}

First, we consider a source of IXs producing radially propagating IXs. For IXs radially propagating from a source with $\psi({\bf r}) \propto e^{i k R}$, where $R=|\mathbf{r} -\mathbf{r}_{\rm s}|$ is the distance to the source, $\mathbf{r}_{\rm s}$ the source location, and $\mathbf{k}=k (\mathbf{r}-\mathbf{r}_{\rm s})/R$ the IX momentum, the interference pattern is given by $I(x,y) = 2 + 2\cos({\bf k} \delta {\bf x} +q_ty)$,  see Fig.~3a. This expression allows estimating the IX momentum $\bf k$ in the vicinity of an LBS source from the measured interference pattern as described in Ref.~\cite{Leonard2018}.

Now, we consider two sources of IXs, each producing radially propagating IXs. For simplicity, to outline the origin of the phase singularities, we consider in Fig.~3b-f the sources producing IX fluxes of equal strength and with equal absolute value of the IX momentum $k$. For the two sources of IXs, the interference pattern is given by
\begin{equation}
I(x,y) \sim \frac{1}{R_1}
\cos ({\bf k}_1\cdot \delta \mathbf{x} + q_t y) +
\frac{1}{R_2} \cos ({\bf k}_2 \delta {\bf x} + q_t y),
\end{equation}
where $R_{i}=|\mathbf{r}-\mathbf{r}_{{\rm s},i}|$ and
${\bf k}_i=k (\mathbf{r}-\mathbf{r}_{{\rm s},i})/R_i$
are the distance to the source and momentum for IXs propagating from source $i$ with $i = 1, 2$. In this expression, $I(x,y)$ intensities produced by source 1 and source 2 are added. The following assumptions are made: (i) each source generates a coherent flux of IXs propagating ballistically from the source with no dissipation, (ii) there is no coherence between the IX fluxes propagating from source 1 and source 2. These assumptions are in accord with the experiment: (i) the long coherence length for the condensate of IXs produced by an LBS source~\cite{High2012} supports the long-range ballistic IX propagation, (ii) the IX creation from optically generated holes and electrically generated electrons in LBS sources~\cite{Yang2015} suggests the lack of coherence between different LBS sources. (We note parenthetically that similar phase singularities in the interference pattern can appear also for the case of coherence between the sources.)

Let us first consider the case of the two sources separated along the shift $\delta\mathbf{x}$ (Fig.~3b). On the bisector line equally separated from two sources Eq.~(1) gives
\begin{equation}
I \sim \cos (q_t y) \cos (k \delta x \cos \gamma),
\end{equation}
where $\gamma$ is the angle between $\mathbf{r}-\mathbf{r}_{{\rm s}1}$ and $\delta \mathbf {x}$, see the case $\beta = 0$ in Fig.~3d schematic. In Eq.~(2), the first factor oscillates along $y$, and the second factor produces the beating pattern. When
\begin{equation}
k \delta x \cos \gamma = \left( n + \frac{1}{2} \right) \pi
\end{equation}
with $n$ an integer, the 2nd factor in Eq.~2 changes sign, this phase slip creates the phase singularity in  interference pattern. Numerical simulations shown in Fig.~3b,e confirm that the dislocations in interference pattern are indeed located at angle $\gamma = \arccos \left[ (k \delta x)^{-1} \pi/2 \right]$, in agreement with Eq.~3. The pairs of phase singularities also appear in simulations at other beating nodes when $\cos (k \delta x \cos \gamma)$ changes sign, i.e. at
$\gamma = \arccos \left[ (k \delta x)^{-1} \left( n + \frac{1}{2} \right) \pi \right]$.

For the region close to the bisector line, i.e. for $x^\prime/R\ll 1$ with $x^\prime$ the distance of the location $(x,y)$ to the bisector, and $R=(R_1+R_2)/2$, Eq.~(1) gives
\begin{eqnarray}
I &\sim& \frac{2}{R} \cos \Big(q_t y +\frac{k x'}{R}\sin^2\gamma\Big)
\cos \Big(k\delta x\cos\gamma \Big) \nonumber \\
&+&\frac{2x' \cos\gamma}{R^2} \sin q_t y \sin \Big( k\delta x\cos\gamma  \Big)
+I_{\rm bg},
\label{eq:interference}
\end{eqnarray}
where $I_{\rm bg}$ is the background. If we move slightly leftward ($x^\prime<0$), or, rightward ($x^\prime>0$) away from the phase slipping point on the bisector line, the 2nd term in Eq.~\ref{eq:interference} dominates leading to the bifurcation of the interference pattern.

For the case of nonzero angle $\beta$ between the line connecting two IX sources and $\bf x$, the phase slip along the bisector line equally separated from the two sources appears at angle $\gamma$ given by $k \delta x \cos \gamma \cos \beta = \pi/2$ (more generally $k \delta x \cos \gamma \cos \beta = \left( n + \frac{1}{2} \right) \pi$), where $\gamma$ is the angle between the line connecting the sources and the direction from the source to the singularity point~(Fig.~3d). Numerical simulations (Fig.~3c,f) confirm that the dislocations in interference pattern are indeed located at angle $\gamma = \arccos \left[ (k \delta x \cos \beta)^{-1} \pi/2 \right]$.

The simulations in Fig.~3b,c reproduce the "isolated" dislocations observed in the experiment (Fig.~1a) and outline their origin: the observed interference dislocations originate from converging of the condensate matter waves propagating from different sources. Here the "isolated" dislocations mean the dislocations separated by a significantly larger distance than the shift-interferometry shift $\delta x$. In contrast, dislocations separated by $\delta x$ can be associated with vortices, or polarization vortices, or half-vortices as decibed above (Fig.~2).

Most close to the observed phase singularities in the IX condensate interference patterns are, probably, wavefront dislocations~\cite{Nye1974} studied in various systems of waves ranging from matter waves~\cite{Berry1980} to light~\cite{Baranova1981}. Those wavefront dislocations are expected whenever limited trains of waves travel in different directions and interfere~\cite{Nye1974}. We note however that along with similarities, there are also substantial differences between the observed phase singularities in the IX condensate interference patterns and wavefront dislocations. For instance, the former do no require spatial modulations of the wave (limited train waves) as the latter. Other peculiarities of the interference dislocations are outlined in their description above.

In the experimental system, there are several LBS sources of IXs with different strength (Fig.~4a). Three strong sources are clearly seen due to the pronounced phase domains at the source locations (Fig.~4a). These phase domains are associated with the Pancharatnam-Berry phase and are described in Ref.~\cite{Leonard2018}. Two weaker sources can also be seen in Fig.~4a. The right- and left-oriented dislocations in the interference pattern in Fig.~4a are marked by red and green circles, respectively. All LBS sources participate in the formation of these dislocations. A stronger contribution to the upper pair of right- and left-oriented dislocations is given by the two upper strong sources, a stronger contribution to the lower pair of right- and left-oriented dislocations is given by the two lower strong sources.

Simulated interference pattern for IX condensates radially propagating from five sources with $I(x, y) \sim \sum_{i=1}^{5} P_i/R_i \cos ({\bf k}_i \delta {\bf x} + q_t y)$ is presented in Fig.~4b. These simulations are similar to the simulations for two sources following Eq.~1 and shown in Fig.~3b,c. $R_i$ and ${\bf k}_i$ are the distance to the source and momentum for IXs propagating from source $i$. The values of ${\bf k}_i$ are estimated from the position of the interference fringes in the source vicinity following the method presented in Ref.~\cite{Leonard2018}. The source powers $P_i$ are estimated from the IX emission intensities in the source region. The sources in the simulations (Fig.~4b) are positioned at the locations of the LBS sources in the experiment (Fig.~4a). The simulations in Fig.~4b qualitatively reproduce the phase singularities marked by red and green circles observed in the experiment in Fig.~4a.

All observed phase singularities disappear above the IX condensation temperature where the interference patterns become trivial with continuous interference fringes showing neither phase jumps no dislocations. This is consistent with the requirement of a long-range ballistic IX propagation outlined above. A long-range ballistic IX propagation is provided by the long coherence length in the condensate.

In this paragraph, we briefly discuss possible future studies. The simulations use a model of radial ballistic IX propagation from the sources. This model reproduces the observed singularities in the interference pattern and reveals their origin. Making the model closer to the experiment by taking into account various experimental factors such as repulsive IX interaction, finite IX lifetime, etc. forms the subject for future works. Furthermore, this work addresses the singularities for converging condensates generated by different sources. A possibility to create similar singularities using a single source and splitting and converging the condensate fluxes generated by this source also forms the subject for future works. In this direction, an interesting issue to address is the possibility of appearance of the singularities due to splitting and converging the fluxes in random potential energy landscapes. Another interesting issue to address is the possibility to create and control the singularities through the condensate flux control in tailored potential energy landscapes. Tailored IX potential energy landscapes $U(x,y) \propto edF_z(x,y)$ can be formed and controlled by voltage due to the IX built-in electric dipole moment $ed$ [$F_z(x,y)$ is electric field in the $z$ direction, $d$ is the separation between the electron and hole layers].

In summary, we present dislocation-like singularities in interference patterns in a condensate of indirect excitons. We show that the observed dislocations in interference patterns are not associated with conventional phase defects: neither with vortices, nor with polarization vortices, nor with half-vortices, nor with skyrmions, nor with half-skyrmions. We present the origin of these new singularities in condensate interference patterns: the observed interference dislocations originate from converging of the condensate matter waves propagating from different sources.

\medskip\noindent {\bf Acknowledgments}
We thank Misha Fogler for discussions. These studies were supported by DOE Office of Basic Energy Sciences under award DE-FG02-07ER46449 and by NSF Grant No.~1640173 and NERC, a subsidiary of SRC, through the SRC-NRI Center for Excitonic Devices.

\medskip\noindent {\bf Methods}
The $n-i-n$ GaAs/AlGaAs CQW heterostructure was grown by molecular beam epitaxy. The $i$ region consists of a single pair of 8-nm GaAs QWs separated by a 4-nm Al$_{0.33}$Ga$_{0.67}$As barrier and surrounded by 200-nm Al$_{0.33}$Ga$_{0.67}$As layers. The $n$ layers are Si-doped GaAs with Si concentration $5 \cdot 10^{17}$~cm$^{-3}$. The indirect regime where IXs form the ground state is realized by the voltage applied between $n$ layers. The small in-plane disorder in the CQW is indicated by the emission linewidth of 1~meV. IXs cool to temperatures within $\sim 50$~mK of the lattice temperature \cite{Butov2001}, which was lowered to 100~mK in an optical dilution refrigerator. This cools IXs well below the temperature of quantum degeneracy, which is in the range of a few kelvin for typical IX density $10^{10}$~cm$^{-2}$~\cite{Butov2001}. The photoexcitation is provided by a 633~nm HeNe laser, more than 400~meV above the energy of IXs and farther than 80~$\mu$m away from the studied region, IX coherence is not induced by photoexcitation and forms spontaneously. LBS are sources of cold IXs due to their separation from the laser excitation spot. The shift in the shift-interferometry experiments $\delta x = 2$~$\mu$m. This shift is suitable for these experiments because it is smaller than the coherence length, smaller than the characteristic sizes of the features in the interference patterns, and larger than the 1.4~$\mu$m spatial resolution in the experiment. The simulations are performed for the same shift. The path lengths of the arms of the MZ interferometer are equal. After the interferometer, the emission is filtered by an interference filter of linewidth $\pm 5$~nm adjusted to the IX emission wavelength $\sim 800$~nm. The filtered signal is focused to produce an image, which is measured by a liquid-nitrogen cooled CCD. The interference pattern is given by $I_{\rm interf} = (I - I_1 - I_2)/(2\sqrt{I_1 I_2})$, where $I_1$ is IX emission intensity for arm 1 open, $I_2$ for interferometer arm 2 open, and $I$ for both arms open.


\begin{references}

\bibitem{Volovik2003}
G.E. Volovik, The Universe in a He-Droplet (Oxford University Press, 2003).

\bibitem{Matthews1999}
M.R. Matthews, B.P. Anderson, P.C. Haljan, D.S. Hall, C.E. Wieman, E.A. Cornell, Vortices in a Bose-Einstein Condensate, {\it Phys. Rev. Lett.} {\bf 83}, 2498 (1999).

\bibitem{Inouye2001}
S. Inouye, S. Gupta, T. Rosenband, A.P. Chikkatur, A. G{\"o}rlitz, T.L. Gustavson, A.E. Leanhardt, D.E. Pritchard, W. Ketterle, Observation of Vortex Phase Singularities in Bose-Einstein Condensates, {\it Phys. Rev. Lett.} {\bf 87}, 080402 (2001).

\bibitem{Hadzibabic2006}
Z. Hadzibabic, P. Kr{\"u}ger, M. Cheneau, B. Battelier, J. Dalibard, Berezinskii-Kosterlitz-Thouless crossover in a trapped atomic gas, {\it Nature} {\bf 441}, 1118 (2006).

\bibitem{Khawaja2001}
U.Al. Khawaja, H. Stoof, Skyrmions in a ferromagnetic Bose-Einstein condensate, {\it Nature} {\bf 411}, 918 (2001).

\bibitem{Kamchatnov2008}
A.M. Kamchatnov, L.P. Pitaevskii, Stabilization of Solitons Generated by a Supersonic Flow of Bose-Einstein Condensate Past an Obstacle, {\it Phys. Rev Lett.} {\bf 100}, 160402 (2008).

\bibitem{Wu2008}
Congjun Wu, Ian Mondragon Shem, Exciton condensation with spontaneous timereversal symmetry breaking, arXiv:0809.3532v1; Cong-Jun Wu, Ian Mondragon-Shem, Xiang-Fa Zhou, {\it Chin. Phys. Lett.} {\bf 28}, 097102 (2011).

\bibitem{Wang2010}
Chunji Wang, Chao Gao, Chao-Ming Jian, Hui Zhai, Spin-Orbit Coupled Spinor Bose-Einstein Condensates, {\it Phys. Rev. Lett.} {\bf 105}, 160403 (2010).

\bibitem{Nowik-Boltyk2012}
P. Nowik-Boltyk, O. Dzyapko, V.E. Demidov, N.G. Berloff, S.O. Demokritov, Spatially non-uniform ground state and quantized vortices in a two-component Bose-Einstein condensate of magnons, {\it Sci. Rep.} {\bf 2}, 482 (2012).

\bibitem{Rubo2007}
Y.G. Rubo, Half Vortices in Exciton Polariton Condensates, {\it Phys. Rev. Lett.} {\bf 99}, 106401 (2007).

\bibitem{Lagoudakis2008}
K.G. Lagoudakis, M. Wouters, M. Richard, A. Baas, I. Carusotto, R. Andr{\'e}, Le Si Dang, B. Deveaud-Pl{\'e}dran, Quantized vortices in an exciton-polariton condensate, {\it Nature Physics} {\bf 4}, 706 (2008).

\bibitem{Lagoudakis2009}
K.G. Lagoudakis, T. Ostatnick{\'y}, A.V. Kavokin, Y.G. Rubo, R. Andr{\'e}, B. Deveaud-Pl{\'e}dran, Observation of Half-Quantum Vortices in an Exciton-Polariton Condensate, {\it Science} {\bf 326}, 974 (2009)

\bibitem{Roumpos2011}
G. Roumpos, M.D. Fraser, A. L{\"o}ffler, S. H{\"o}fling, A. Forchel, Y. Yamamoto, Single vortex-antivortex pair in an exciton-polariton condensate, {\it Nature Physics} {\bf 7}, 129 (2011).

\bibitem{Amo2011}
A. Amo, S. Pigeon, D. Sanvitto, V.G. Sala, R. Hivet, I. Carusotto, F. Pisanello, G. Lem{\' e}nager, R. Houdr{\' e}, E. Giacobino, C. Ciuti, A. Bramati, Polariton Superfluids Reveal Quantum Hydrodynamic Solitons, {\it Science} {\bf 332}, 1167 (2011).

\bibitem{Grosso2011}
G. Grosso, G. Nardin, F. Morier-Genoud, Y. L{\' e}ger, B. Deveaud-Pl{\' e}dran, Soliton Instabilities and Vortex Street Formation in a Polariton Quantum Fluid, {\it Phys. Rev. Lett.} {\bf 107}, 245301 (2011).

\bibitem{Flayac2011}
H. Flayac, D.D. Solnyshkov, G. Malpuech, Oblique half-solitons and their generation in exciton-polariton condensates, {\it Phys. Rev. B} {\bf 83}, 193305 (2011).

\bibitem{Hivet2012}
R. Hivet, H. Flayac, D.D. Solnyshkov, D. Tanese, T. Boulier, D. Andreoli, E. Giacobino, J. Bloch, A. Bramati, G. Malpuech, A. Amo, Half-solitons in a polariton quantum fluid behave like magnetic monopoles, {\it Nature Physics} {\bf 8}, 724 (2012).

\bibitem{Flayac2013}
H. Flayac, D.D. Solnyshkov, I.A. Shelykh, G. Malpuech, Transmutation of Skyrmions to Half-Solitons Driven by the Nonlinear Optical Spin Hall Effect, {\it Phys. Rev. Lett.} {\bf 110}, 016404 (2013).

\bibitem{Bardyn2015}
C.-E. Bardyn, T. Karzig, G. Refael, T.C.H. Liew, Topological polaritons and excitons in garden-variety systems,  {\it Phys. Rev. B} {\bf 91}, 161413(R) (2015).

\bibitem{High2012}
A.A. High, J.R. Leonard, A.T. Hammack, M.M. Fogler, L.V. Butov, A.V. Kavokin, K.L. Campman, A.C. Gossard, Spontaneous coherence in a cold exciton gas, {\it Nature} {\bf 483}, 584 (2012).

\bibitem{Matuszewski2012}
M. Matuszewski, T.C.H. Liew, Y.G. Rubo, A.V. Kavokin, Spin-orbit coupling and the topology of gases of spin-degenerate cold excitons in photoexcited GaAs-AlGaAs quantum wells, {\it Phys. Rev. B} {\bf 86}, 115321 (2012).

\bibitem{Kyriienko2012}
O. Kyriienko, E.B. Magnusson, I.A. Shelykh, Spin dynamics of cold exciton condensates, {\it Phys. Rev. B} {\bf 86}, 115324 (2012).

\bibitem{High2013}
A.A. High, A.T. Hammack, J.R. Leonard, Sen Yang, L.V. Butov, T. Ostatnick{\' y}, M. Vladimirova, A.V. Kavokin, T.C.H. Liew, K.L. Campman, A.C. Gossard, Spin currents in a coherent exciton gas. {\it Phys. Rev. Lett.} {\bf 110}, 246403 (2013).

\bibitem{Kavokin2013}
A.V. Kavokin, M. Vladimirova, B. Jouault, T.C.H. Liew, J.R. Leonard, L.V. Butov, Ballistic spin transport in exciton gases, {\it Phys. Rev. B} {\bf 88}, 195309 (2013).

\bibitem{Vishnevsky2013}
D.V. Vishnevsky, H. Flayac, A.V. Nalitov, D.D. Solnyshkov, N.A. Gippius, G. Malpuech, Skyrmion Formation and Optical Spin-Hall Effect in an Expanding Coherent Cloud of Indirect Excitons, {\it Phys. Rev. Lett.} {\bf 110}, 246404 (2013).

\bibitem{Sigurdsson2014}
H. Sigurdsson, T.C.H. Liew, O. Kyriienko, I.A. Shelykh, Vortices in spinor cold exciton condensates with spin-orbit interaction, {\it Phys. Rev. B} {\bf 89}, 035302 (2014).

\bibitem{Li2016}
Yi Li, X. Zhou, C. Wu, Three-dimensional quaternionic condensations, Hopf invariants, and skyrmion lattices with synthetic spin-orbit coupling. {\it Phys. Rev. A} {\bf 93}, 033628 (2016).

\bibitem{Leonard2018}
J.R. Leonard, A.A. High, A.T. Hammack, M.M. Fogler, L.V. Butov, K.L. Campman, A.C. Gossard, Pancharatnam-Berry phase in condensate of indirect excitons, {\it Nature Commun.} {\bf 9}, 2158 (2018).

\bibitem{Lozovik1976}
Yu.E. Lozovik, V.I. Yudson, A new mechanism for superconductivity: pairing between spatially separated electrons and holes. {\it Sov. Phys. JETP} {\bf 44}, 389 (1976).

\bibitem{Yang2015}
Sen Yang, L.V. Butov, B.D. Simons, K.L. Campman, A.C. Gossard, Fluctuation and commensurability effect of exciton density wave, {\it Phys. Rev. B} {\bf 91}, 245302 (2015).

\bibitem{Scheuer1999}
J. Scheuer, M. Orenstein, Optical Vortices Crystals: Spontaneous Generation in Nonlinear Semiconductor Microcavities, {\it Science} {\bf 285}, 230 (1999).

\bibitem{Nye1974}
J.F. Nye, M.V. Berry, Dislocations in wave trains, {\it Proc. R. Soc. Lond. A} {\bf 336}, 165 (1974).

\bibitem{Berry1980}
M.V Berry, R.G. Chambers, M.D. Large, C. Upstill, J.C. Walmsley, Wavefront dislocations in the Aharonov-Bohm effect and its water wave analogue, {\it Eur. J. Phys.} {\bf 1}, 154 (1980).

\bibitem{Baranova1981}
N.B. Baranova, B.Ya. Zel'dovich, A.V. Mamaev, N.F. Pilipetskii, V.V. Shkukov, Dislocations of the wavefront of a speckle-inhomogeneous field (theory and experiment), {\it JETP Lett.} {\bf 33}, 195 (1981).

\bibitem{Butov2001}
L.V. Butov, A.L. Ivanov, A. Imamoglu, P.B. Littlewood, A.A. Shashkin, V.T. Dolgopolov, K.L. Campman, A.C. Gossard, Stimulated Scattering of Indirect Excitons in Coupled Quantum Wells: Signature of a Degenerate Bose-Gas of Excitons, {\it Phys. Rev. Lett.} {\bf 86}, 5608 (2001).

\end{references}
\end{document}